\documentclass{ws-ijmpa}
\usepackage{amsmath,amssymb,graphicx,cite}
\usepackage{dcolumn}% Align table columns on decimal point
\usepackage{subfigure}
\usepackage{bm}% bold math
\usepackage{soul}
\usepackage{hyperref}
\graphicspath{{Figures/}}
\begin{document}

\title{A note on supergravity inflation in braneworld}

\author{Mudassar Sabir}
\address{Institute of Geophysics and Geomatics, China University of Geosciences, Wuhan, Hubei 430074, China  
	%sabir@cug.edu.cn
}
\author{Waqas Ahmed}
\address{School of Physics, Nankai University, Tianjin 300350, China \\
	%waqasmit@nankai.edu.cn
}
\author{Yungui Gong}
\address{School of Physics, Huazhong University of Science and Technology, Wuhan, Hubei 430074, China \\
	%yggong@hust.edu.cn
}
\author{Shan Hu}
\address{Department of Physics, Faculty of Physics and Electronic Sciences, Hubei University, Wuhan, Hubei 430062, China\\
	%hushan@hubu.edu.cn
}
\author{Tianjun Li}
\address{CAS Key Laboratory of Theoretical Physics, Institute of Theoretical Physics, Chinese Academy of Sciences, Beijing 100190, China \\
	School of Physical Sciences, University of Chinese Academy of Sciences, Beijing, China \\
	%tli@itp.ac.cn
}
\author{Lina Wu}
\address{School of Sciences, Xi'an Technological University, Xi'an, Shanxi 710021, China \\
	%wulina@xatu.edu.cn
}

\maketitle

%\begin{history}
%	\received{Day Month Year}
%	\revised{Day Month Year}
%\end{history}

\begin{abstract}
We discuss supergravity inflation in braneworld cosmology for the class of potentials $V(\phi)=\alpha \phi^n\rm{exp}(-\beta^m \phi^m)$ with $m=1,~2$. These minimal SUGRA models evade the $\eta$ problem due to a broken shift symmetry and can easily accommodate the observational constraints. In the high energy regime $V/\lambda\gg 1$, the numerical predictions and approximate analytic formulas are given for the scalar spectral index $n_s$ and tensor-to-scalar ratio $r$. The models with smaller $n$ are preferred while the models with larger $n$ are out of the $2\sigma$ region. Remarkably, the $\rho^2/\lambda$ correction to the energy density in Friedmann equation results in sub-Planckian inflaton excursions $\Delta\phi <1$.
\end{abstract}

%\preprint{1905.03033}

\section{Introduction}\label{sec:Intro}

It is well-known that our universe may have experienced accelerated expansion, {\it i.e.},
cosmic inflation~\cite{Starobinsky:1980te, Guth:1980zm, Linde:1981mu, Albrecht:1982wi}, at a very early stage of its evolution as evident from the temperature fluctuations and observations of polarization spectrum of the cosmic microwave background radiation. It provides a convincing explanation for the observed large scale homogeneity and isotropy of the universe.
 As we know, the energy scale of inflation related to tensor-to-scalar ratio $r$~\cite{Lyth:1984yz} is given by
\begin{equation}
V^{1/4}\simeq \left(\frac{r}{0.01}\right)^{1/4}\times10^{16}\rm ~GeV.
\end{equation}
From Planck 2018 combined with the BICEP2/Keck Array BK14 data \cite{Akrami:2018odb,Aghanim:2018eyx}, this ratio between the amplitudes of the tensor mode and scalar mode of CMB $r=A_t/A_s$ is limited: $r_{0.05} \le 0.06~(95\%~{\rm CL})$.
Thus, the slow low-roll inflation may have taken place at the unification scale in Grand Unified Theory (GUT) or below \cite{Wu:2016fzp, Yi:2018gse, Ahmed:2018jlv} and is assumed to be described by the effective field theories (EFTs) \cite{Cheung:2007st}.
Such EFTs can only be trusted if they can be successfully embedded in a theory of quantum gravity. Models in standard general relativity suffer from the problem of super-Planckian field excursions.

As we know, the inflaton excursion $\Delta\phi$ and tensor-to-scalar ratio $r$ are related via
 the Lyth bound~\cite{Lyth:1996im,Gao:2014pca} as follows
\begin{eqnarray}\label{Eq:lythbound}
\Delta\phi &\simeq& N \sqrt{\frac{r}{8}}~,~\,
\end{eqnarray}
where $N$ is the number of $e$-folds before the end of inflation. To avoid super-Planckian inflaton excursions
and satisfy the flatness requirement of $N\geq50$, we obtain $r \leq 0.003$. This upper bound of $r$ is
on the edge of observability at the future LiteBIRD experiment, which is designed for the detection
of B-mode polarization pattern embedded in the Cosmic Microwave Background anisotropies~\cite{Suzuki:2018cuy}.

Supersymmetry provides the most promising extension for the Standard Model of particle physics. However, it is still regarded as a global symmetry. In the context of AdS/CFT \cite{Maldacena:1997re, Witten:1998qj}, it was argued in Ref. \cite{Harlow:2018tng} that no global symmetries can exist in a theory of quantum gravity. It is thus believed that supersymmetry must be a local symmetry, {\it i.e.} supergravity. The theory of supergravity is a natural framework for inflation model building~\cite{Freedman:1976xh, Deser:1976eh}.

On the other hand, in the braneworld scenario, the brane tension parameter $\lambda$ induces a $\rho^2$ correction to the energy density in Friedmann equation \cite{Shiromizu:1999wj,Csaki:1999jh,Maartens:1999hf,Binetruy:1999hy,Binetruy:1999ut,Maartens:2010ar,Mukerji:2011yv}, which affects the behavior of inflaton $\phi$ and results in sub-Planckian inflaton excursions. In the high energy regime $\rho\gg\lambda$ or $V\gg\lambda$, the modified slow-roll parameters are suppressed by the factor $V/\lambda$~\cite{Bento:2001hu,Bento:2002kp,Paul:2003jx,Sami:2003my,Papantonopoulos:2004bm,Bento:2006sr,Safsafi:2012rx,Ferricha-Alami:2015eqn,Salamate:2019oyl}, and then the Lyth bound can be violated. Thus, the problem of super-Planckian field excursion can be solved~\cite{Lin:2018kjm,Brahma:2018hrd,Sabir:2019xwk, Sabir:2019wel,Mohammadi:2020ctd,Sakhi:2020gcr}. Interestingly, the slow-roll conditions can be satisfied even by the steep inflaton potential~\cite{Ida:1999ui,Cline:1999ts,Copeland:2000hn,Myung:2001sp,Palma:2005wm}.

In this paper, we consider a class of inflationary models with the potential $V(\phi)=\alpha \phi^n\rm{exp}(-\beta^m \phi^m)$
that can be obtained in supergravity inflation with a small shift symmetry breaking term in K\"ahler potential and has
relatively large tensor-to-scalar ratio~\cite{Li:2013nfa}.
We embed such inflationary models into the braneworld scenario, and show that the large tensor-to-scalar ratio can be
smaller and within the 1 $\sigma$ (95\% C.L.) bounds of the Planck data, as well as
the inflaton excursions can be sub-Planckian.
Therefore, such kind of the models not only meets the distance criteria \cite{Obied:2018sgi,Ooguri:2018wrx},
but also satisfies all the current experimental constraints from TT, TE, EE$+$lowE$+$lensing$+$BK14$+$BAO data
in Ref.~\cite{Akrami:2018odb,Aghanim:2018eyx}.

\section{Brane inflation in supergravity with broken shift symmetry}\label{sec:SUGRA}
We will consider a class of inflationary models with the scalar potential as the following form
\begin{equation}
    V=\alpha\phi^n \exp(-\beta^m \phi^m), \quad m=1,2; \quad n=1,~3/2,~2~  \label{eq:pot}
\end{equation}
which can be derived from supergravity with a small shift symmetry breaking term
in K\"ahler potential \cite{Li:2013nfa},  in which the exponential term $e^{-\beta^m\phi^m}$ is crucial to fit the Planck results
  \cite{Enqvist:1985yc,Li:2013moa,Mazumdar:2014bna}. To be more specific, one can derive this potential by starting with a shift symmetry preserved term $-\frac{1}{2}(\Phi-\bar{\Phi})^2$ and a shift symmetry breaking term $-b^m(\Phi^m+\bar{\Phi}^m)$ in K\"ahler potential. It is possible that the scalar potential could be affected by the bulk gravity and the shift symmetry breaking term could lead to a large tensor-to-scalar ratio and cause the trans-Planckian inflaton excurions \cite{Kawasaki:2000yn,Yamaguchi:2000vm,Kawasaki:2001as,Brax:2005jv,Kallosh:2010ug,Kallosh:2010xz,Li:2013nfa,Lazarides:2015xia}. Our models have a valuable merit in phenomenology that they can be embedded into the five-dimensional brane scenario. In the next section, we will show that after being embedded into the five-dimensional brane scenario, the flatter potential in the inflaton direction $\phi=\sqrt{2}{\rm Re}(\Phi)$ gives a consistent CMB observation and sub-Planckian inflaton excursions.
In the following, we will discuss more details about such embeddings.

In braneworld cosmology, our 4-dimensional world is a 3-brane embedded in a higher-dimensional bulk. The Friedmann equation with a $\rho^2/\lambda$ correction is \cite{Shiromizu:1999wj,Csaki:1999jh,Maartens:1999hf,Binetruy:1999hy,Binetruy:1999ut,Maartens:2010ar,Mukerji:2011yv}
\begin{equation}
H^2=\frac{1}{3M_{\rm P}^2}\rho \left( 1+\frac{\rho}{2 \lambda} \right),\label{eq:Friedmann}
\end{equation}
where $\rho$ is the observable 3-brane energy density, and $\lambda$ is
the brane tension parameter defined as follows
\begin{equation}
\lambda \equiv \frac{3}{4\pi} \frac{M_5^6}{M_4^2}~.
\end{equation}
where $M_4$ and $M_5$ are the four-dimensional and five-dimensional Planck scales, respectively . And the reduced Planck mass is $M_{\rm P}=M_4/\sqrt{8\pi}$.
From the nucleosynthesis, we obtain $\lambda \gtrsim (1 \mbox{ MeV})^4 \sim (10^{-21})^4 M_{\rm P}^4$. Because the theory must be reduced to the Newtonian gravity on scales larger than 1 mm, we get a stronger constraint $\lambda \gtrsim 5 \times 10^{-53} M_{\rm P}^4$, i.e. $M_5 \gtrsim 10^5$ TeV.
Also, the standard Friedman equation in four dimensions is recovered at the low energy limit or
large $\lambda$ limit with $\rho/\lambda \rightarrow 0$.

The slow-roll parameters in brane scenario are modified into \cite{Maartens:1999hf}
\begin{eqnarray}
\epsilon &=& \frac{1}{2} \left( \frac{V^\prime}{YV} \right)^2 \left( 1+\frac{V}{\lambda} \right), \label{epsilon} \\
\eta &=&  \frac{V^{\prime\prime}}{YV} ~,~\, \label{eta}
\end{eqnarray}
where
\begin{eqnarray}
Y&= & 1+\frac{V}{2\lambda} ~.~\,
\end{eqnarray}
And the number of $e$-foldings are calculated as
\begin{equation}
N = -\int_{\phi_*}^{\phi_{e}}\frac{YV}{V'}  d{\phi}~.\label{eq:efolds}
\end{equation}
In the low energy regime with $V/\lambda\ll 1$, the traditional definitions for the standard cosmology are recovered. However, in the high energy regime with $V/\lambda\gg 1$, the slow-roll parameters
$\epsilon$ and $\eta$ are further suppressed by a factor $V/\lambda$. Thus,
the slow-roll conditions can be satisfied even by the steep inflaton potential,
and we shall consider the high energy regime with $V/\lambda \gg 1$ in this paper.

We shall consider the Randall-Sundrum model II,
whose amplitudes for the scalar and tensor power spectrum are \cite{Randall:1999vf,Langlois:2000ns}
\begin{eqnarray}
A_s = \frac{1}{12 \pi^2}\frac{(YV)^3}{V'^{2}}~,~
A_t =\frac{2}{3\pi^2}  YV  F^2~, \label{eq:PsPt}
\end{eqnarray}
where
\begin{equation}
F^2=\left[\sqrt{1+x^2} -x^2 \sinh^{-1}\left(\frac{1}{x}\right)\right]^{-1}~,\label{eq:F2}
\end{equation}
and
\begin{align}
\label{eq:x}
x\equiv \left(\frac{3 \,H^2}{4 \pi\,\lambda}\right)^{1/2} = \left[\frac{2YV} {\lambda}\right]^{1/2}~.
\end{align}

In terms of the slow-roll parameters, the  scalar spectral index $n_s$
and the tensor-to-scalar ratio $r$ are given by~\cite{Bento:2008yx}
\begin{equation}
\begin{split}
n_{s} &= 1 -6\epsilon + 2\eta~,\\
r &=\frac{A_t}{A_s}=8\left(\frac{V'}{V}\right)^2\left(\frac{F}{Y}\right)^2~.
\end{split}
\end{equation}
Therefore, compared to the four-dimensional inflationary models, we find that $n_s$ is the same, while $r$ is different.
In the low energy limit with $V/\lambda\ll 1$ and  $F^2\approx 1$, we get the traditional result $r=16\epsilon$.
However, in the high energy limit with $V/\lambda\gg 1$ and $F^2\approx 3\,V/2\,\lambda$,
we obtain the modified result $r=24\epsilon$, and the modified  amplitude of the scalar power spectrum as follows
\begin{eqnarray}
A_s\simeq\frac{V}{12\pi^2\epsilon}\left(\frac{V}{2\lambda}\right)^2.
\end{eqnarray}
Therefore, we cannot fix the energy scale of inflation and the brane tension
via the observational values of $r$ and $A_s$.

\section{Numerical results}\label{sec:Numerical}

\begin{figure}[t]
	\centering
	\includegraphics[height=5.0cm]{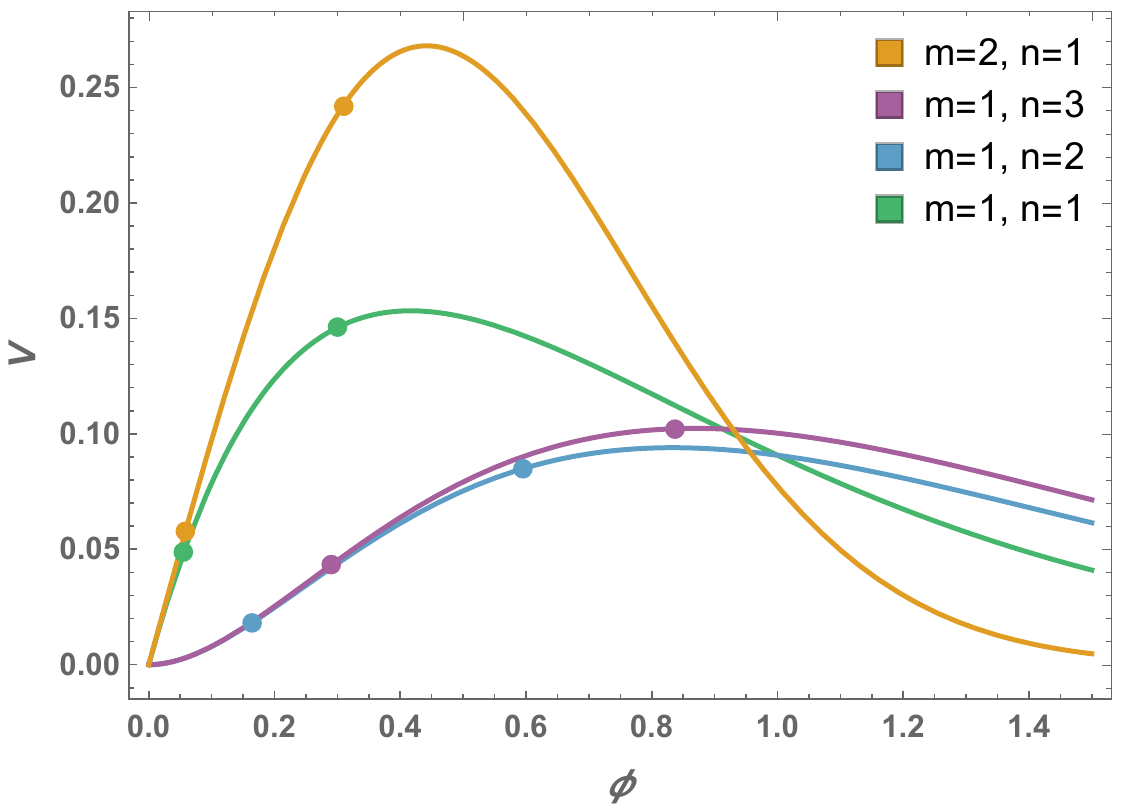}
	\caption{Potentials for models with $m=1,~2$. The parameters $\beta$ and $\alpha$ are chosen to be consistent with the Planck observations $n_s=0.965$ and $A_{s}=2.10\times10^{-9}$. The points represent $\phi_*$ and $\phi_e$.}\label{Fig:pot}
\end{figure}

We numerically calculate the inflationary predictions and check the consistency conditions for both $m=1$ and $m=2$ models.
The selected inflation path is between the origin and the turning point $\phi_t=\beta^{-1}(m^{-1}n)^\frac{1}{m}$ as indicated in Figure \ref{Fig:pot}.

\subsection{Predictions for $n_{s}-r$}

\begin{figure}[htb]
	\centering
	\subfigure{\includegraphics[height=4.3cm]{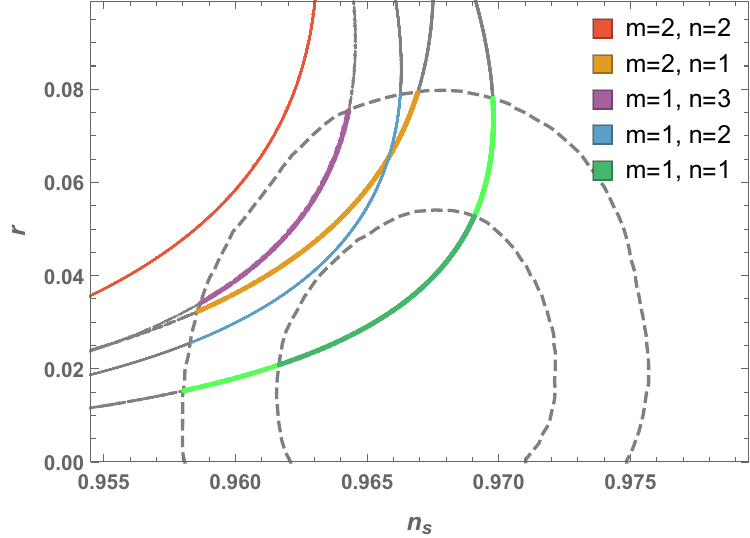}}\quad
	\subfigure{\includegraphics[height=4.3cm]{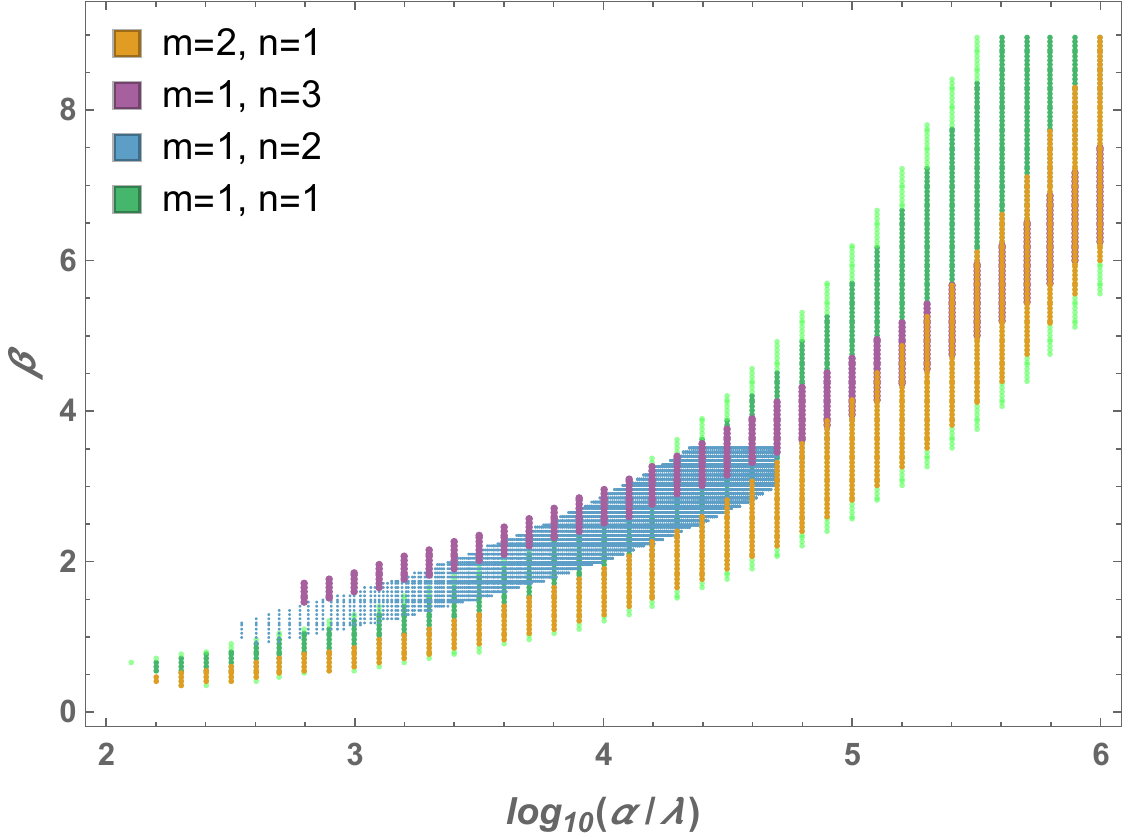}}\\
	\caption{$n_{s}-r$ plane for $m=1,2$ models with $N=60$ and $\Delta\phi<1$. The dashed contours come from Planck 2018 data \cite{Akrami:2018odb,Aghanim:2018eyx}. In the right panel, $1\sigma$ and $2\sigma$ constrained parametric space for $\log_{10}{(\alpha/\lambda)}$ and $\beta$ is shown with the corresponding colors.}\label{Fig:ns_r}
\end{figure}

Figure~\ref{Fig:ns_r} gives the $1\sigma$ and $2\sigma$ results of scalar spectral index $n_s$ and tensor-to-scalar ratio $r$ for both types of models. We set the number of $e$-folds to be $N=60$ and let the two parameters $\alpha/\lambda$ and $\beta$ vary independently with the sub-Planckian field excursions condition $\Delta\phi<1$ imposed.

It can be seen from Figure~\ref{Fig:ns_r} that the model parameters $\alpha/\lambda$ and $\beta$ are correlated. The value of $\beta$ increases as $\alpha/\lambda$ increases. Hence, we just need to determine the lower limits for the parameters $(\alpha/\lambda;~\beta)$. For the models $m=1,~n=1$ and $m=1,~n=2$, the lower limits are $(150;~0.65)$ and $(350;~1)$, for the models $m=2,~n=1$ and $m=1,~n=2$, they are $(150;~0.45)$ and $(900;~0.8)$ in reduced Planck units.

For all the models inflation ends at $\epsilon=1$ since $|\eta|$ is always smaller than $\epsilon$ at the inflation exit point. For models with $m=1$, the slow-roll parameter $\eta$ is
\begin{eqnarray}
\eta &=& \left( \frac{V^{\prime\prime}}{V} \right)\left( \frac{1}{1+\frac{V}{2 \lambda}} \right)\nonumber\\
&=&\left(\dfrac{n}{\phi}\left(\dfrac{n-1}{\phi}-2\beta\right)+\beta^2\right) \left( \dfrac{1}{1+\frac{V}{2 \lambda}} \right).
\end{eqnarray}

\subsection{Analytic approximations for $n_s$ and $r$}

For models under the consideration, $\lambda/\alpha \ll 1$, we can expand the number of $e$-foldings $N$, the spectral index $n_s$, and the tensor-to-scalar ratio $r$ up to first order in $\lambda/\alpha$ as
\begin{gather}
N ~\approx~  \frac{\alpha}{2 \beta ^3 \lambda}\left(e^{-\beta\phi}(\beta  \phi+2)+\frac{\text{Ei}(1-\beta \phi)}{e}\right)\Bigl|_{\phi_e}^{\phi_*}+{\cal O}\left(\lambda/\alpha\right)^2~,\label{eq:Napp}\\
n_s(\phi_*) ~\approx~ 1-\frac{4\beta ^3\lambda}{\alpha} \frac{ e^{\beta  \phi_* } \left( 2 \beta ^2 \phi_* ^2- 4 \beta  \phi_* +3\right)}{(\beta \phi_*) ^3}+{\cal O}\left(\lambda/\alpha\right)^2~,\label{eq:nsapp}\\
r(\phi_*)  ~\approx~  \frac{48 \beta ^3\lambda }{\alpha} \frac{  e^{\beta  \phi_* } (1-\beta \phi_*)^2}{(\beta \phi_*)^3}+{\cal O}\left(\lambda/\alpha\right)^2~,\label{eq:rapp}
\end{gather}
where Ei$(x)$ is the exponential integral.
It is apparent from Eq.~\eqref{eq:Napp} that $\beta\phi_*$ is a
function of $(\beta ^3 \lambda N)/\alpha$.
Similarly from Eqs.~\eqref{eq:nsapp} and \eqref{eq:rapp} both $n_s$ and $r$
depend on $(\beta ^3 \lambda )/\alpha$ for a fixed $N$.
This explains our numerical results in Figure~\ref{Fig:ns_r}
where we obtained a single curve for each model with two varying parameters $\alpha/\lambda$ and $\beta$, which, effectively combine into a single one $\alpha/(\beta ^3 \lambda)$. Furthermore, for $N=60$, we should have $\alpha/(\beta ^3 \lambda)\gtrsim 10^2$ to ensure that the spectral index remains less than unity. Hence the parameters $\alpha/\lambda$ and $\beta$ are bounded from below as can be seen from the right panel in Figure~\ref{Fig:ns_r}.

\subsection{$ A_{s}$ and the range of $ \alpha $, $ \lambda $}

\begin{figure}[t]
	\centering
	\subfigure[]{\includegraphics[height=4.3cm]{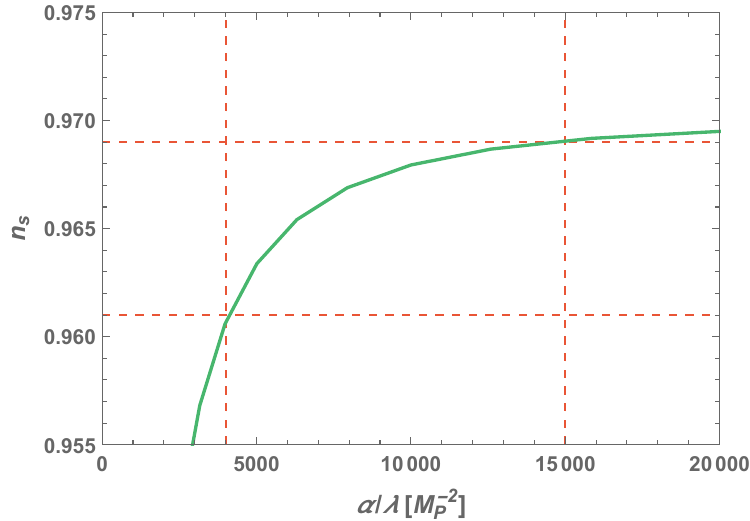}}\quad
	\subfigure[]{\includegraphics[height=4.3cm]{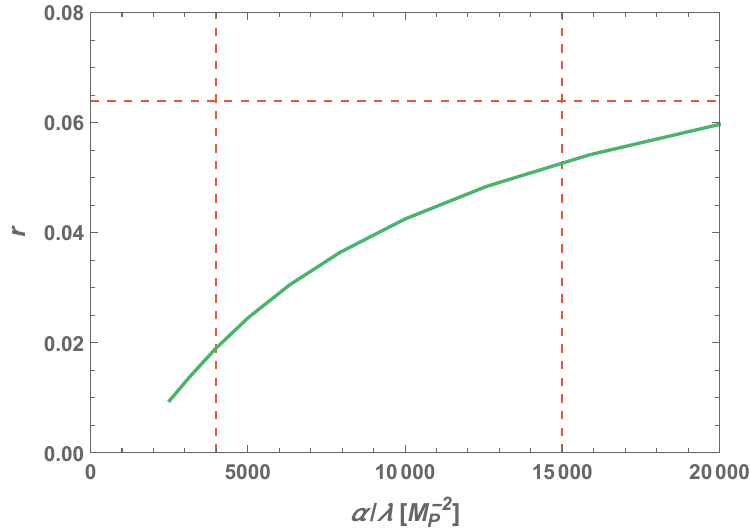}}\\
	\caption{Result for $m=1,~n=1$ model, $A=\alpha/\lambda$ is the variable with $\beta=2$ and $N=60$. The vertical lines locate at $A_1=4000~M_{\rm P}^{-2}  $ and $A_2=15000~M_{\rm P}^{-2}$. The horizontal lines in (a) correspond to the 68\% CL bound of $n_s=0.9649\pm 0.0042$~\cite{Akrami:2018odb,Aghanim:2018eyx}. 95\% CL upper bound on the tensor-to-scalar ratio from Planck 2018 combined with the BICEP2/Keck Array BK14 data is $r_{0.05}<0.06$ \cite{Akrami:2018odb,Aghanim:2018eyx} which is given as the horizontal line in (b).}\label{Fig:A}
\end{figure}

In order to realize the observed power spectrum $A_{s}=2.10\times10^{-9}$, either $\lambda$ or $\alpha$ should be adjusted accordingly. In Figure~\ref{Fig:A}, the range of $A=\alpha/\lambda$ in $m=1,~n=1$ model with $\beta=2$ is given.
Comparing with the 2018 Planck results \cite{Akrami:2018odb,Aghanim:2018eyx}, when $A\in[4000, 15000]~M_{\rm P}^{-2}$, the spectral index of scalar perturbations $n_s$ locates in the 68\% CL bound of $n_s=0.9649\pm 0.0042$. In this parametric region, the tensor-to-scalar ratio $r$ is significantly smaller than its upper bound $r_{0.05}<0.06$. To realize  $A_{s}=2.10\times10^{-9}$, we need $\alpha\in [1.5\times10^{-14},~5.5\times10^{-14}]~M_{\rm P}^3$ and $\lambda\in [1.1\times10^{-18},~1.4\times10^{-17}]~M_{\rm P}^5$. Hence for $\beta\sim 2$ and $\alpha/\lambda\sim10^{4}~M_{\rm P}^{-2}$, the typical values are $\alpha\sim10^{-14}~M_{\rm P}^3$ and $\lambda\sim 10^{-18}~M_{\rm P}^5$.

Figure~\ref{Fig:pot} shows the potentials for models with $m=1,2$. The inflation ends when the slow-roll condition $\epsilon=1$ satisfied. The points in Figure~\ref{Fig:pot} correspond to  $\phi_*$ and $\phi_e$. Here, the parameters $\alpha$ and $\beta$ are chosen to be consistent with the Planck observations. As we can see, $\beta$ is $\mathcal{O}(1)$ to ensure $60$ $e$-foldings. The parameters and results are also given in Table~\ref{Tab:pot}.

\begin{table}[!h]
	\begin{tabular}{ccccccc}
		\hline
		(m,n)&$\beta$&$\alpha(10^{-14}M_{\rm P}^3)$&$n_s$&$r$&$\phi_*(M_{\rm P})$&$\phi_e(M_{\rm P})$\\\hline
		(1,1)&2.39&2.38&0.9649&0.0289&0.3003&0.0555\\
		(1,2)&2.37&21.63&0.9649&0.0541&0.5954&0.1657\\
		(1,3)&2.33&39.98&0.9646&0.0900&0.8373&0.2906\\
		(2,1)&1.56&1.04&0.9650&0.0591&0.3087&0.0579\\\hline
	\end{tabular}
	\caption{ $\alpha/\lambda$ is $10^4$, and $e$-foldings $N=60$.}\label{Tab:pot}
\end{table}

\begin{figure}[!h]
	\centering
	\subfigure{\includegraphics[height=4.3cm]{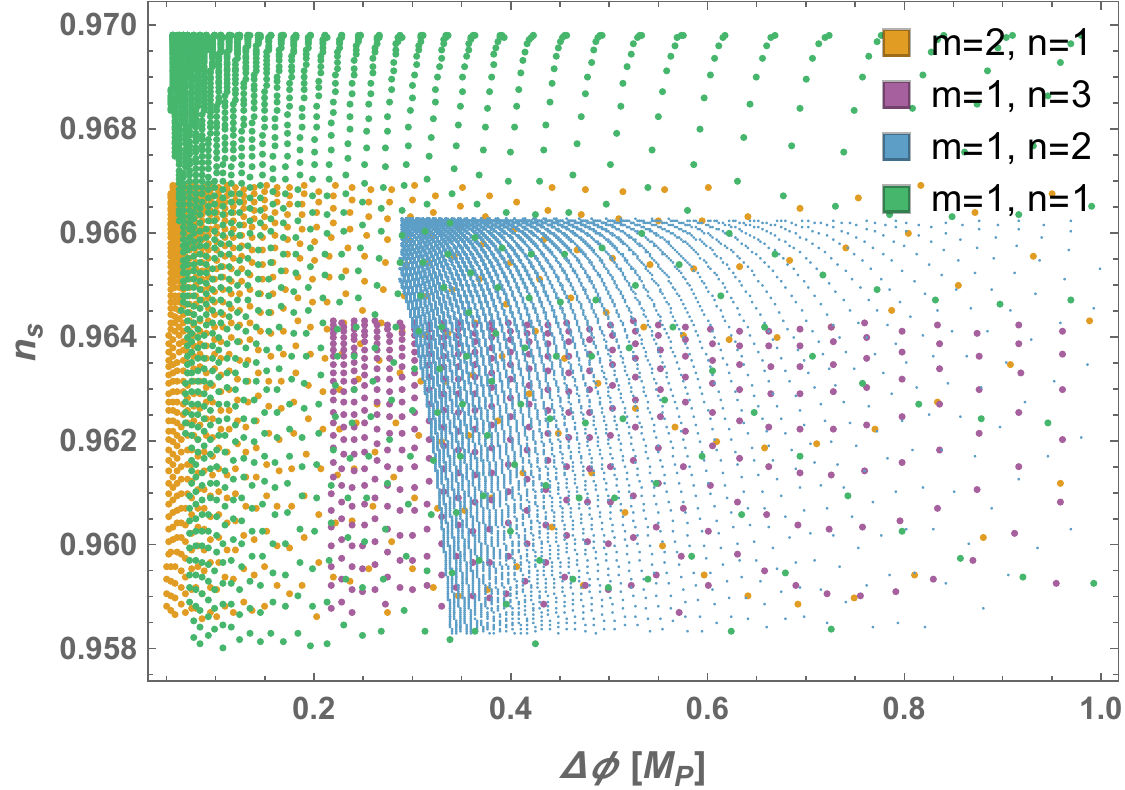}}\quad
	\subfigure{\includegraphics[height=4.3cm]{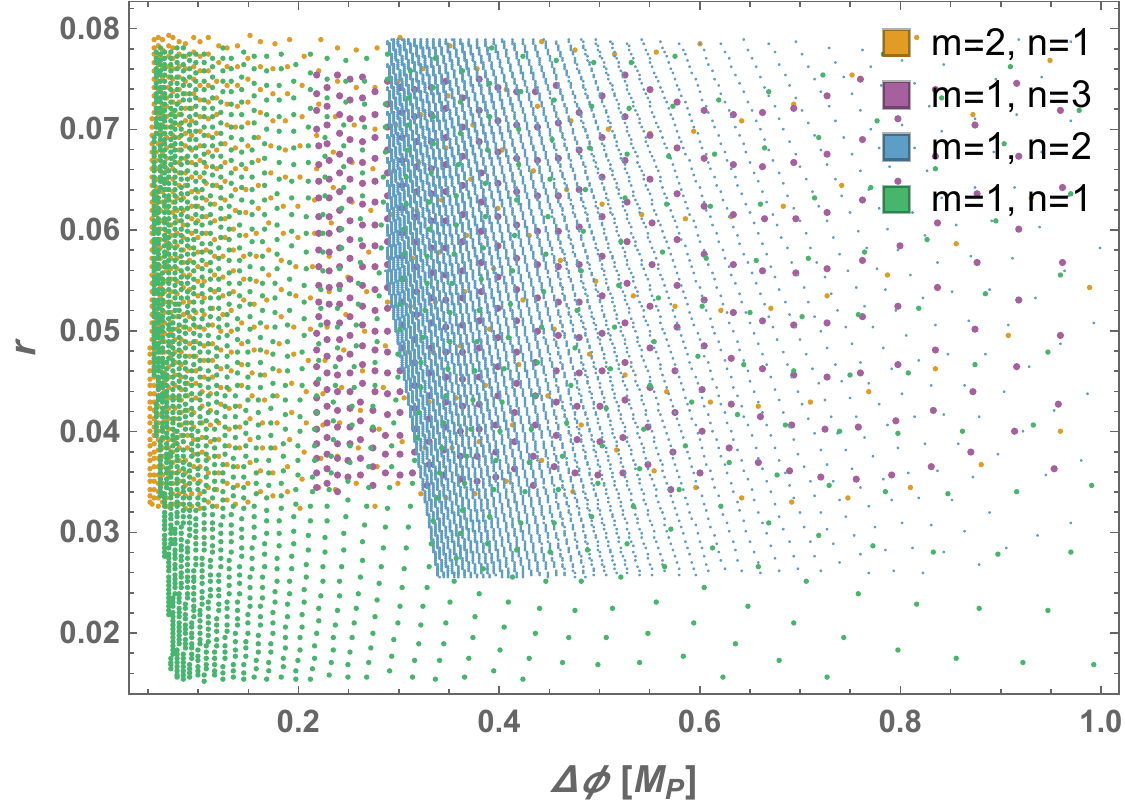}}\\
	\caption{Left panel shows the plot of $n_{s}$ and field excursion $\Delta\phi$ while the right panel shows the plot of $r$ versus $\Delta\phi$ for $m=1,2$ models with $N=60$ $e$-foldings.}\label{Fig:dphi_ns_r}
\end{figure}

In Figure \ref{Fig:dphi_ns_r} we plot the $n_s$ and $r$ against the field excursion $\Delta\phi$ with $N=60$ $e$-folds for various $m=1,\,2$ models. It can be seen clearly that the field excursion remains sub-Planckian as to avoid the quantum gravity corrections that may break the effective field description due to quantum gravity corrections as advocated in distance conjecture \cite{Obied:2018sgi,Ooguri:2018wrx}.

\section{Conclusions}\label{sec:Conclusion}

We have analyzed a class of potentials in braneworld inflation numerically and analytically.
The class of potentials $V(\phi)=\alpha \phi^n\rm{exp}(-\beta^m \phi^m)$ were obtained in supergravity with a small shift symmetry breaking
term in the K\"ahler potential.

In the high energy regime $V/\lambda\gg 1$ where the brane correction is important, the observables $n_s$ and $r$ depend on a single parameter $\beta^3\lambda/\alpha$ although there are three independent parameters $\lambda$, $\beta$
and $\lambda$ in the model. This novel and interesting result is apparent from numerical analysis and shown
explicitly with analytical approximation for the model with $m = n = 1$. The sub-Planckian field
inflation models with $m = 1$ and $m = 2$ accommodate the Planck 2018 observational constraints.
The inflation models with smaller $n$ are preferred while the models with larger $n$ are out of the
2$\sigma$ bound in $n_s-r$ plane. In particular, the model with $m = 1$ and $n = 1$ is consistent with
the Planck observations at the 1$\sigma$ confidence level.

In braneworld the slow-roll conditions can even be satisfied for otherwise steep potentials
and the amount of inflation between any two points in the potential is greater than
the one obtained in standard cosmology. The corresponding field excursions also stay sub-Planckian.
The primordial gravitational waves associated to the obtained value of $r\sim 10^{-2}$ might be detected with the future LiteBIRD satellite.

\section*{Acknowledgments}
This research was supported by the Projects 11475238, 11605049, 11647601, 11875062, and 11875136 supported by the National Natural Science Foundation of China, by the Major Program of the National Natural Science Foundation of China under Grant No. 11690021, by the Key Research Program of Frontier Science (CAS), by the Program 2020JQ-804 supported by Natural Science Basic Research Plan in Shanxi Province of China, and by the Program 20JK0685 funded by Shanxi Provincial Education Department.

\bibliographystyle{ws-ijmpa}
%\bibliography{References}

\end{document}